\documentclass[%
 reprint,
 amsmath,amssymb,
 aps,
]{revtex4-1}

\usepackage{graphicx}
\usepackage{dcolumn}
\usepackage{bm}
\usepackage[Symbol]{upgreek}

\begin{document}


\title{Fiber-optic two-way quantum time transfer with frequency-entangled pulses}

\author{Feiyan Hou$^{1}$}
\author{Runai Quan$^{1}$}
\author{Ruifang Dong$^{1}$}
\email{dongruifang@ntsc.ac.cn}
\author{Xiao Xiang$^{1}$}
\author{Baihong Li$^{1,2}$}
\author{Tao Liu$^{1}$}
\author{Xiaoyan Yang$^{3,4}$}
\author{Hao Li$^{3,4}$}
\author{Lixing You$^{3,4}$}
\email{lxyou@mail.sim.ac.cn}
\author{Zhen Wang$^{3,4}$}
\author{Shougang Zhang$^{1}$}

\affiliation{$^1$ Key Laboratory of Time and Frequency Primary Standards, National Time Service Center, Chinese Academy of Sciences, Xi'an, 710600, China}
\affiliation{$^2$ College of Sciences, Xi'an University of Science and Technology, Xi'an, 710054, China}
\affiliation{$^3$  State Key Laboratory of Functional Materials for Informatics, Shanghai Institute of Microsystem and Information Technology, Chinese Academy of Sciences, Shanghai 200050, China}
\affiliation{$^4$ Center for Excellence in Superconducting Electronics, Chinese Academy of Sciences, Shanghai, 200050, China}
\date{\today}

\begin{abstract}
High-precision time transfer is of fundamental interest in physics and metrology. Quantum time transfer technologies that use frequency-entangled pulses and their coincidence detection have  been proposed, offering potential enhancements in precision and better guarantees of security. 
In this paper, we describe a fiber-optic two-way quantum time transfer experiment. Using quantum nonlocal dispersion cancellation, time transfer over a $20$-km fiber link achieves a time deviation of $922$ fs over $5$ s and $45$ fs over $40960$ s. The time transfer accuracy as a function of fiber lengths from $15$ m to $20$ km is also investigated, and an uncertainty of $2.46$ ps in standard deviation is observed. In comparison with its classical counterparts, the fiber-optic two-way quantum time transfer setup shows appreciable improvement, and further enhancements could be obtained by using new event timers with sub-picosecond precision and single-photon detectors with lower timing jitter for optimized coincidence detection. Combined with its security advantages, the femtosecond-scale two-way quantum time transfer is expected to have numerous applications in high-precision middle-haul synchronization systems.
\begin{description}
\item[PACS numbers]
42.50.Dv, 42.50.Lc, 42.62.Eh
\end{description}
\end{abstract}

\pacs{to be added}
\maketitle


\section{\label{sec:level1}Introduction}

The ability to precisely compare and synchronize distant clocks is increasingly important for contemporary space geodesy, high-resolution radio-astronomy, modern particle physics, navigation and positioning, and for almost every type of precision measurement. Benefitting from the low loss, high reliability, and high stability of optical fibers, time transfer over fibers (TTOF) offers potentially superior performance over its satellite-based counterparts. Over fiber lengths of hundreds of kilometers \cite{krehlik2012,sliwczynski2013,lopez2013}, accuracies of tens of picoseconds have been reported, marking a significant improvement over satellite-based two-way time transfer techniques \cite{Rabindran2008,Jiang2017}. To improve the time transfer performance, various fiber-based time transfer techniques have been investigated \cite{krehlik2012,sliwczynski2013,lopez2013,piester2009,rost2012,lopez2015,wang2012,yin2014,raupach2014,chen2015,krehlik2017,lessing2017}. The use of mode-locked lasers as sources has recently attracted interest, and sub-femtosecond time transfer has been demonstrated over dispersion-compensated kilometer-scale optical fibers \cite{peng2013,xin2014,ning2014}. However, all time transfer implementations over longer fiber links have been carried out using direct amplitude-modulation either on continuous-wave lasers \cite{piester2009,rost2012,lopez2015,wang2012,yin2014,raupach2014,chen2015,krehlik2017} or a frequency comb \cite{lessing2017}. By compensating the fluctuations of the propagation delay of the optical fibers, the stability of the two-way transfer approach remains at the sub-picosecond level up to averaging times of hours. On the other hand, secure time transfer is critical to widespread technologies and infrastructures that rely on distributed precision time, such as military use of navigation system\cite{jafarnia2012}, financial networks\cite{angel2014}.
Though two-way transfer enables the detection of man-in-the-middle (MITM) delay attacks, classical techniques are susceptible to interference by malicious parties, which adversely affects the overall functionality of the dependent applications \cite{narula2018}. 

To accurately and securely distribute time information among distant clocks, there is a compelling need for fundamentally new methods. The single-photon avalanche detector (SPAD) has many advantages for measuring the time-of-flight of optical pulses. Apart from its ability to detect extremely low-power signals, high precision, and elimination of most systematic errors, another distinct feature is its preservation of data traffic security. SPAD has been applied to space laser time transfer experiments, achieving a time stability of $1$ ps over $1000$ s and $10$ ps over $1$ day \cite{exertier2014}, and an accuracy better than $100$ ps \cite{samain2015}. Together with the use of frequency-entangled photon-pairs as the timing source, quantum-enhanced time-transfer is highly anticipated \cite{narula2018,giovannetti2011,giovannetti2001a,giovannetti2001b,bahder2004,wang2016,valencia2004,quan2016,quan2019}. Furthermore, secure time synchronization is guaranteed by the complementarity principle of quantum mechanics \cite{giovannetti2002,lee2014,Lamas-Linares2018}. Due to the quantum effect that arises from the strong correlation between photons originating from the entanglement, an adversary must be able to perform quantum non-demolition measurements of the presence of a single photon with high probability to compromise the security, which is a serious technological barrier for any would-be adversary \cite{Lamas-Linares2018}. Based on a Bell inequality test,  the security of the system can be thus ensured \cite{lee2014,Lamas-Linares2018}. 

Despite its potential high precision and cryptographically secure nature, the superiority of quantum time transfer is somewhat underappreciated because of the relatively low photon rate, as well as the dispersion deterioration of the pulse propagating through the fiber. Fortunately, with regard to the dispersion in the fiber, nonlocal dispersion cancellation associated with frequency-entangled photons have been proposed in 1992 \cite{franson1992} and experimentally demonstrated subsequently through local detection \cite{baek2009,O2011,maclean2018}. Since the dispersion experienced by the signal photons can be canceled nonlocally by the idler photons after experiencing an opposite dispersion, the two-photon coincidence can be recovered from the dispersion deterioration. In this paper, we demonstrate that, combined with the unique property of quantum nonlocal dispersion cancellation associated with frequency-entangled pulses, a quantum two-way time transfer over fiber (Q-TWTTOF) scheme can outperform the analogous classical schemes. The performance of this scheme has been simulated based on the quantum model. Through theoretical analysis, we show that the  time stability in terms of standard deviation is determined by the spectral bandwidth of the entangled photon pairs and the dispersive broadening induced by the fiber. Exploiting nonlocal dispersion compensation \cite{franson1992,baek2009,O2011} can thus improve the transfer stability. Q-TWTTOF experiments have been conducted on a $20$-km fiber coiling, in which a dispersion compensation fiber (DCF)  of length $2.49$ km has been inserted into the idler arm for the nonlocal dispersion cancellation. Experimental results show that the time transfer stabilities in terms of time deviation (TDEV) reach $922$ fs over an averaging time of $5$ s and $45$ fs over $40960$ s.  
Presently, performance is mainly restricted by the limited acquisition rate of the event timing system and the timing jitter of the single-photon detector and the event timers (ETs). Compared with classical analogues, the time transfer stability has shown appreciable improvement and could be further enhanced by using new ETs with sub-picosecond precision \cite{Panek2013} and single-photon detectors with lower timing jitter \cite{Wu2017}.  The dependence on fiber length has been investigated by measuring the clock differences as a function of fiber lengths from $15$ m to $20$ km, resulting in a variation of $2.46$ ps in standard deviation (SD). This reveals that our system has superior time transfer symmetry ($ \le {\rm{0}}{\rm{.12}}$ps $km^{-1}$) against symmetric channel delay attacks \cite{lee2018}. Together with its inherent security advantage, the two-way quantum time transfer method has the potential to be very useful for highly accurate and secure time transfer over modest distances. 

The remainder of this paper is organized as follows. Section II provides a brief description of the Q-TWTTOF scheme. Section III presents its theoretical derivation. Section IV describes the experimental setup and section V presents results and analysis. Section VI contains our conclusions.
\section{Schematic description}
\begin{figure}[ht]
\centerline{
	\includegraphics[width=0.48\textwidth]{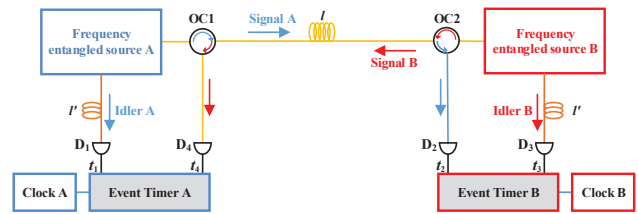}}
	\caption{\label{fig1} Sketch of the Q-TWTTOF between two clocks at sites A and B. A and B each have a frequency-entangled  photon-pair source, a pair of SPADs, and an ET referenced to its local time scale. The idler photon of the pair is detected locally after passing through a piece of DCF of length $l'$, while the signal photon is sent through a single-mode fiber of length $l$ to be detected on the remote side. The times of arrival for all detected photons are recorded at each side with respect to the local clock. OC: optical circulator, which is used for bidirectional fiber transmission, D1$-$D4: SPADs for single-photon detection.}
\end{figure}

The scheme for realizing the two-way quantum time transfer between two clocks at separate sites (A and B) that are interconnected via a fiber link is sketched in Fig. \ref{fig1}. Each site has a frequency-entangled photon-pair source, a pair of SPADs, and an ET referenced to its local time scale. For the entangled source generated at site A, the signal photons travel from A to B through a fiber of length $l$ while the idler photons are held locally at site A. To compensate for the dispersion experienced by signal photons in the fiber link, a piece of DCF of length $l'$ is inserted into the idler path. 
With the help of the SPADs (D1 \& D2), the transmitted signal and idler photons are detected. Using the ETs, the times of arrival for the detected photons are then recorded at each site with respect to the local clock. Assume that the arrival times of signal A and the idler A photons are recorded as $\{ {t_2^{\left( j \right)}} \}$ and $\{ {t_1^{\left( j \right)}} \}$ , respectively, where $j = 1...n$  denotes a series of time-tagged sequences. By applying a cross-correlation algorithm to them \cite{quan}, the coincidence histogram of the time difference between ${t_2}$ and ${t_1}$ can be constructed. Through Gaussian fitting of the coincidence distribution, the registration time difference ${t_2} - {t_1}$ with respect to the maximum “coincidences” is obtained. Assume the time difference between clock A and clock B is ${t_0}$ . Then, it is deduced that ${t_2} - {t_1} = l/{\upsilon _{g,A}} - l'/{\upsilon '_{g,A}} - {t_0}$, where ${\upsilon _g}$ and ${\upsilon '_g}$ are the group velocities in the propagation path and DCF. For the frequency-entangled source at site B, a similar procedure is carried out. The signal photons travel from B to A through the same fiber link and the idler photons are held at site B after traveling through the same type of DCF with the same length as that at site A. From the recorded arrival times  $\{ {t_4^{\left( j \right)}} \}$ and $\{ {t_3^{\left( j \right)}} \}$ , the time difference ${t_4} - {t_3}$ can be extracted; this is expressed as ${t_4} - {t_3} = l/{\upsilon _{g,B}} - l'/{\upsilon '_{g,B}} + {t_0}$. If ${\upsilon _{g,A}} = {\upsilon _{g,B}}$ and ${\upsilon '_{g,A}} = {\upsilon '_{g,B}}$ are satisfied, the time difference between the two clocks is then given by ${t_0} = \tfrac{{\left( {{t_4} - {t_3}} \right) - ({t_2} - {t_1})}}{2}$.

\section{Theoretical analysis}
According to quantum field theory, the probability of coincidentally detecting the above timing events at space-time points $\left( {{D_1},{t_1}} \right),\left( {{D_2},{t_2}} \right),\left( {{D_3},{t_3}} \right),\left( {{D_4},{t_4}} \right)$ is proportional to the fourth-order correlation function of the fields \cite{glauber1963}
\begin{widetext}
\begin{equation}
{G^{\left( 4 \right)}} = \left\langle \Psi  \right|E_1^{\left(  -  \right)}E_2^{\left(  -  \right)}E_3^{\left(  -  \right)}E_4^{\left(  -  \right)}E_4^{\left(  +  \right)}E_3^{\left(  +  \right)}E_2^{\left(  +  \right)}E_1^{\left(  +  \right)}\left| \Psi  \right\rangle,
\end{equation}
\end{widetext}
where $E_j^{\left( \pm \right)}$ refers to the positive and negative components of the electric field at the $\textit{j}$-th detector, which is directly related to the annihilation (creation) operators ${\hat a_j}$ (${\hat a_j}^\dag$).
\begin{eqnarray}
\begin{array}{l}
E_j^{\left(  +  \right)}\left( {{D_j},{t_j}} \right) \propto {{\hat a}_j}\left( {{D_j},{t_j}} \right),\\
E_j^{\left(  -  \right)} = {( {E_j^{\left(  +  \right)}})^\dag }, j = 1,2,3,4,
\end{array}
\end{eqnarray}
where ${\hat a_j}$ denotes the annihilation operators at the $\textit{j}$-th detector.
$\left| \Psi  \right\rangle $ is the state of the input field. Let the states of the two frequency-entangled sources at A and B be given by $\left| {{\Phi _A}} \right\rangle $ and $\left| {{\Theta _B}} \right\rangle $, respectively. $\left| \Psi  \right\rangle $ can then be given by the direct product $\left| \Psi  \right\rangle  = \left| {{\Phi _A}} \right\rangle  \otimes \left| {{\Theta _B}} \right\rangle $. Assuming they are generated through the same degenerate type-II spontaneous parametric down-conversion (SPDC) process via a monochromatic pump, the state functions can be written as
\begin{eqnarray}
\begin{array}{l}
\begin{array}{*{20}{c}}
{\left| {{\Phi _A}} \right\rangle  = \mathop \smallint \nolimits d\Omega f\left( \Omega  \right)\hat a_{s,A}^\dag \left( {{\omega _0} + \Omega } \right)\hat a_{i,A}^\dag \left( {{\omega _0} - \Omega } \right)\left| 0 \right\rangle ,}\\
{\left| {{\Theta _B}} \right\rangle  = \mathop \smallint \nolimits d\Omega g\left( \Omega  \right)\hat a_{s,B}^\dag \left( {{\omega _0} + \Omega } \right)\hat a_{i,B}^\dag \left( {{\omega _0} - \Omega } \right)\left| 0 \right\rangle ,}
\end{array}
\end{array}
\end{eqnarray}

where $\hat a_{s,A\left( B \right)}^\dag $ and $\hat a_{i,A\left( B \right)}^\dag $ are the creation operators for the signal and idler photons of the frequency-entangled source generated at site A (B). $\left| 0 \right\rangle $ represents the vacuum state. Ideally, the signal and idler photons are frequency anticorrelated, with their spectra centered around $\omega_0$, and $f\left( {\rm{\Omega }} \right) = g\left( {\rm{\Omega }} \right) = {\rm{sinc}}\left( {{{DL{\rm{\Omega }}} \mathord{\left/{\vphantom {{DL{\rm{\Omega }}} {\rm{2}}}} \right.\kern-\nulldelimiterspace} {\rm{2}}}} \right)$ denotes the joint spectral amplitude function of the frequency-entangled states. Here, $L$ is the SPDC crystal length and $D$ is the inverse group velocity difference for signal and idler photons crossing the crystal. To simplify the deduction, the above expression can be approximated as a Gaussian function $f\left( {\rm{\Omega }} \right) \approx {e^{ - \gamma {{\left( {{\rm{\Omega }}DL} \right)}^2}}}$ with $\gamma  = 0.04822$.

In terms of the annihilation operators of the signal and idler photons, ${\hat a_j}$ at the $j$-th detector is given by
\begin{eqnarray}
\begin{array}{l}
{{\hat a}_1}\left( {l',{t_1}} \right) = \int d\omega {{\hat a}_{i,A}}\left( \omega  \right){e^{ - i\omega \left( {{t_1} - {t_{A0}}} \right)}}{e^{ik'\left( \omega  \right)l'}},\\
{{\hat a}_2}\left( {l,{t_2}} \right) = \int d\omega {{\hat a}_{s,A}}\left( \omega  \right){e^{ - i\omega ({t_2} - {t_{B0}})}}{e^{ik(\omega )l}},\\
{{\hat a}_3}\left( {l',{t_3}} \right) = \int d\omega {{\hat a}_{i,B}}\left( \omega  \right){e^{ - i\omega ({t_3} - {t_{B0}})}}{e^{ik'(\omega )l'}},\\
{{\hat a}_4}\left( {l,{t_4}} \right) = \int d\omega {{\hat a}_{s,B}}\left( \omega  \right){e^{ - i\omega ({t_4} - {t_{A0}})}}{e^{ik(\omega )l}}.
\end{array}
\end{eqnarray}

where ${t_{A0}}$ and ${t_{B0}}$ refer to the time at which clock A and clock B start; thus, the time difference between the two clocks is denoted as ${t_0} = {t_{A0}} - {t_{B0}}$. $k\left( \omega  \right)$ and $k'\left( \omega  \right)$ describe the propagation constants over the transmission fiber of length $l$ and the DCF of length $l'$, respectively. Their Taylor expansions around the center frequency ${\omega _0}$ until the second-order term are given by
\begin{widetext}
\begin{eqnarray}
\begin{array}{l}
{k\left( \omega  \right) = \frac{{n\left( \omega  \right)\omega }}{c} = {k_0} + {k_1}\left( {\omega  - {\omega _0}} \right) + \frac{1}{{2!}}{k_2}{{\left( {\omega  - {\omega _0}} \right)}^2},}\\
{k'\left( \omega  \right) = \frac{{n'\left( \omega  \right)\omega }}{c} = {k'_0} + {k'_1}\left( {\omega  - {\omega _0}} \right) + \frac{1}{{2!}}{k'_2}{{\left( {\omega  - {\omega _0}} \right)}^2}.}
\end{array}
\end{eqnarray}
\end{widetext}

Substituting Eqs. (2)$-$(5) into Eq. (1), we can rewrite the fourth-order correlation function as \cite{franson1992,baek2009}
\begin{eqnarray}
{G^{(4)}}(\tau ,\tau ') = {G^{(2)}}\left( \tau  \right){G^{(2)}}\left( {\tau '} \right) \propto {e^{ - \left( {\frac{{{\tau ^2} + {{\tau '}^2}}}{{{2\sigma ^2}}}} \right)}},
\end{eqnarray}
where $\tau  = \left( {{t_4} - {t_3}} \right) - {t_0} - {k_1}l + {k'_1}l'$ ,
 $\tau ' = \left( {{t_2} - {t_1}} \right) + {t_0} - {k_1}l + {k'_1}l'$ and $\sigma  = \sqrt {\gamma {D^2}{L^2} + \frac{1}{{\gamma {D^2}{L^2}}}{{\left( {\frac{{{k_2}l + {k'_2}l'}}{2}} \right)}^2}} $ . From this expression, $\left| {{k_2}l + {k'_2}l'} \right | \to 0$  can be approached by ensuring that ${k_2}$ and ${k'_2}$  have opposite signs. The full width at half maximum (FWHM) of the two ${G^{(2)}}$ functions is given by ${\Delta \tau} _{FWHM} = \Delta {{\tau '}_{FWHM}} \simeq 2.355\sigma $. By taking the integral $\iint d\tau d\tau '\left( {\tau  - \tau '} \right){G^{\left( 4 \right)}}\left( {\tau ,\tau '} \right)$, $\left\langle {\tau  - \tau '} \right\rangle  = 0$ can be derived, corresponding to the classical conclusion of $\left\langle {{t_0}} \right\rangle  = \frac{{\left\langle {\left( {{t_4} - {t_3}} \right) - \left( {{t_2} - {t_1}} \right)} \right\rangle }}{2}$. The SD of ${t_0}$ is determined by 
\begin{widetext}
\begin{eqnarray}
\begin{array}{l}
\Delta {t_0}{\rm{ = }}\frac{{\rm{1}}}{{\rm{2}}}\sqrt {{\Delta ^{\rm{2}}}\left( {{t_4} - {t_3}} \right) + {\Delta ^{\rm{2}}}\left( {{t_2} - {t_1}} \right)} , \\
{\Delta ^{\rm{2}}}\left( {{t_4}{\rm{ - }}{t_3}} \right) = \int {d\tau {{\left[ {\left( {{t_4}{\rm{ - }}{t_3}} \right) - \left\langle {{t_4}{\rm{ - }}{t_3}} \right\rangle } \right]}^2}{G^{\left( 2 \right)}}\left( \tau  \right) = } 4{\sigma ^2},\\
{\Delta ^{\rm{2}}}\left( {{t_2}{\rm{ - }}{t_1}} \right) = \int {d\tau '{{\left[ {\left( {{t_2}{\rm{ - }}{t_1}} \right) - \left\langle {{t_2}{\rm{ - }}{t_1}} \right\rangle } \right]}^2}{G^{\left( 2 \right)}}\left( {\tau '} \right) = } 4{\sigma ^2}.
\end{array}
\end{eqnarray}
\end{widetext}
Substituting Eq. (6) into Eq. (7), $\Delta {t_0}{\rm{ = }}\sqrt{2}\sigma $ is deduced. For detectors with perfect time resolution, given a large number $N$ of detected photon pairs, the deviation should be given by
\begin{widetext}
\begin{eqnarray}
{\left\langle {{\rm{\Delta }}{t_0}} \right\rangle _N}{\rm{ = }}\frac{1}{{\sqrt {2N} }}\left\langle {{\rm{\Delta }}{t_0}} \right\rangle  = \frac{1}{{\sqrt {N} }}\sqrt {\gamma {D^2}{L^2} + \frac{1}{{\gamma {D^2}{L^2}}}{{\left( {\frac{{{k_2}l + {{k'}_2}l'}}{2}} \right)}^2}} .
\end{eqnarray}
\end{widetext}

In practice, the jitters of the single-photon detectors and the ETs for tagging the arrival times of the detected photons will contribute an additive detection response term $\Delta {t_{jitter}}$ to $\Delta {t_0}$, thus the observed ${\left\langle {\Delta {t_0}} \right\rangle _N}$ in Eq.(8) should be rewritten as Eq.(9)
\begin{widetext}
	\begin{eqnarray}
	{\left\langle {\Delta {t_0}} \right\rangle _N} = \frac{1}{{\sqrt N }}\sqrt {\left( {\gamma {D^2}{L^2} + \frac{1}{{\gamma {D^2}{L^2}}}{{\left( {\frac{{{k_2}l + {{k'}_2}l'}}{2}} \right)}^2}} \right) + \Delta t_{jitter}^2} .
	\end{eqnarray}
\end{widetext}
In our experimental setup, the FWHM jitters of the employed SNSPDs and the ETs were measured to be about 70 ps. Two type-II periodically poled potassium titanyl phosphate (PPKTP) crystals of length $10$ mm were used to generate the frequency-entangled sources at $1560$ nm. Thus, $DL=2.96$ ${\rm{ ps}}$. Consider that the length of the transmission fiber is $l=20$ ${\rm{ km}}$  with  ${k_2} \sim  - 2.17 \times {10^{ - 26}}$ ${{\rm{s}}^2}{\rm{/m}}$; the length of DCF is $l'=2.49$ ${\rm{ km}}$ with ${k'_2}\sim  1.86 \times {10^{ - 25}}$ ${{\rm{s}}^2}{\rm{/m}}$. Based on Eq. (8), the FWHM ${\Delta \tau _{FWHM}}$ with and without dispersion compensation can be estimated as $53$ ps and $786$ ps, respectively. After including the contribution from the detection response FWHM jitter, the observed FWHM for the two cases are then estimated to be 88 ps and 789 ps. The corresponding SD ${\left\langle {{\rm{\Delta }}{t_0}} \right\rangle _N}$ are then derived as ${{3.72 \times {{10}^{ - 11}}} \mathord{\left/
		{\vphantom {{3.72 \times {{10}^{ - 11}}} {\sqrt N }}} \right.
		\kern-\nulldelimiterspace} {\sqrt N }}$ ${\rm{s}}$ and ${{3.35 \times {{10}^{ - 10}}} \mathord{\left/
		{\vphantom {{3.35 \times {{10}^{ - 10}}} {\sqrt N }}} \right.
		\kern-\nulldelimiterspace} {\sqrt N }}$ ${\rm{s}}$. Clearly, nonlocal dispersion cancellation can improve the time stability by almost one order of magnitude for the same coincidence rate. Furthermore, as nonlocal dispersion cancellation can increase the coincidence rate \cite{baek2009}, the improvement may be more significant.
\section{Experimental setup}
\begin{figure*}
	\centerline{
		\includegraphics[width=0.8\textwidth]{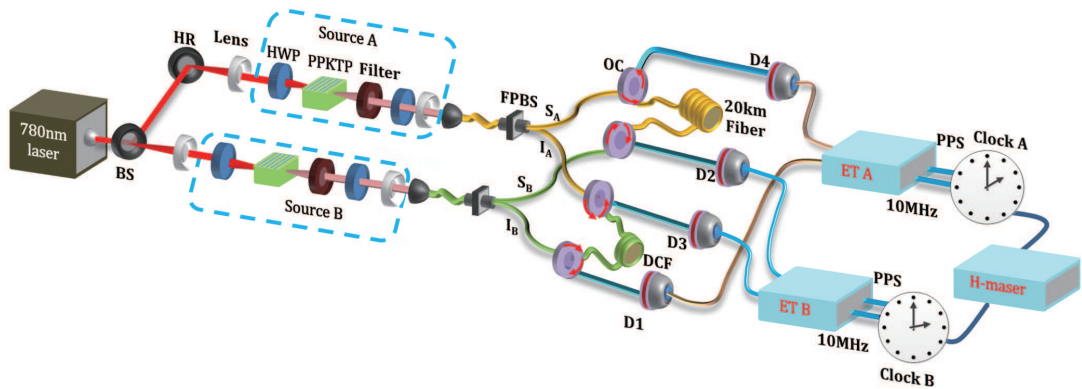}}
	\caption{\label{fig2} Schematic diagram of the experimental setup of the Q-TWTTOF. The 780-nm laser was split into two beams to pump two PPKTP crystals and generate frequency-entangled photon pairs. The collected signal photons A (${S_A}$) and B (${S_B}$) were transmitted in opposite directions over a 20-km fiber coiling, while the idler photons A (${I_A}$) and B (${I_B}$) were transmitted in opposite directions over a 2.49-km DCF. The photons were detected by D1$-$D4, and the arrival times were recorded by ET A and ET B, which were synchronized to clocks A and B, respectively. Clocks A and B used the same H-maser frequency standard. HR: high reflectivity mirror; HWP: half wave plate.}
\end{figure*}
In the experiment, the Q-TWTTOF setup was installed in a common laboratory environment. The ambient temperature of the air-conditioned lab was measured periodically varying within a range of 18.7 $^{\circ}$C to 20.3 $^{\circ}$C. A schematic diagram is shown in Fig.\ref{fig2}. To generate the frequency-entangled photon-pair sources, a quasi-monochromatic $780$-nm laser was produced by the cavity-based frequency doubling process \cite{Hou2016} and then split into two beams by a $50:50$ beam splitter (BS). Each beam was focused into a type-II PPKTP crystal of length 10 mm and a poling period of $46.146$ $\mu$m. After filtering out the residual $780$-nm pump, the output orthogonally polarized frequency-entangled photon pairs were coupled into the fiber polarization beam splitters (FPBS) for spatial separation and subsequent fiber transmission. The transmission fiber link was a $20$-km fiber coiling, and the utilized DCF was $2.49$-km long. The frequency-entangled photons were then detected by superconducting nanowire single-photon detectors (SNSPDs) with an efficiency of 50\%. Commercial ET (Eventech Ltd, A033-ET) were used to record the arrival times of the detected photons. There are two input ports for each ET. The two ports of ET A recorded time-tag sequences $\left\{ {t_1^{\left( j \right)}} \right\}$ and $\left\{ {t_4^{\left( j \right)}} \right\}$, while $\left\{ {t_2^{\left( j \right)}} \right\}$ and $\left\{ {t_3^{\left( j \right)}} \right\}$ were recorded by ET B. The data rate limitation of the ETs, meant that the input signal rate of each port was set to around $6$ kHz. As the data acquisition time was $5$ s, each sequence contained approximately $30000$ time tags. To evaluate the time transfer performance, both ETs were referenced to a common time scale based on the laboratory's own H-maser frequency standard. To further investigate the transfer accuracy in terms of the dependence on fiber length, the absolute time differences were measured for fiber lengths ranging from $15$ m to $20$ km. By utilizing the existing fiber rolls (1 km, 2 km, 3 km, 10 km, 20 km) and pigtails (12 cm, 28 cm, 25 cm) in our lab, we built the experiment for different fiber lengths. The lengths consisted of 15m, 15.12 m, 15.4 m, 15.65 m, 16 m, 16.12 m, 16.37 m, 16.5 m, 16.75 m, 17 m, 1 km, 2 km, 3 km, 10 km, 11 km, 12 km, 13 km, and 20 km. In the experiment, we actually chose three sets of DCF connections. When the fiber was shorter than 3 km, there was no DCF. When the fiber was ranging from 10 to 13 km, the length of the DCF was set to 1.245 km. For the 20 km fiber transmission, 2.49 km-long DCF was applied.

\section{Results and analysis}

\begin{figure}[htb]
	\centerline{
		\includegraphics[width=0.55\textwidth]{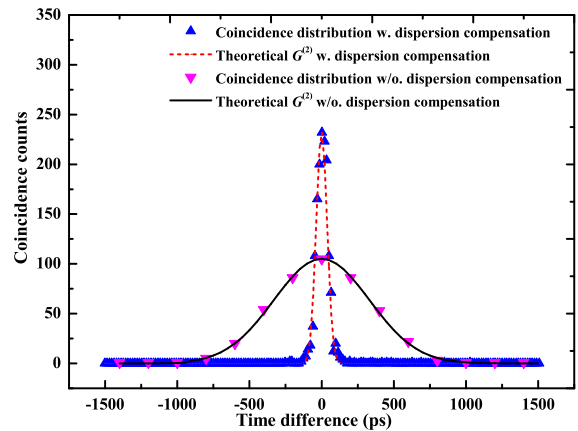}}
	\caption{\label{fig3} Experimentally recovered coincidence distributions from the approximately $30000$ tagged time sequences $\{ {t_2^{\left( j \right)}} \}$ and $\{ {t_1^{\left( j \right)}} \}$, and the theoretically simulated ${G^{(2)}}$ functions with and without DCF in the idler arm. }
\end{figure}

The temporal coincidence distribution histograms recovered from the approximately $30000$ tagged time sequences $\left\{ {t_2^{\left( j \right)}} \right\}$  and $\left\{ {t_1^{\left( j \right)}} \right\}$ are shown in Fig.\ref{fig3}. With the DCF in the idler arm (blue up-triangles), the FWHM is narrowed to $88$ ps with a total coincidence of about $1468$. Without the DCF, the histogram (magenta down-triangles) exhibits a Gaussian-fitted width of about $789$ ps and a total coincidence about $430$. Based on the experimental parameters, the theoretical ${G^{(2)}}$ distributions for the two cases are shown as red dashed and black curves. There is very good agreement between the experiments and the theory. Therefore, both nonlocal coincidence measurement and nonlocal dispersion cancellation have been successfully achieved. Based on Eq. (9), the SDs of ${t_0}$  can be estimated to be about $1.0$ ps and $16.2$ ps for the two cases.

\begin{figure}[htb]
	\centerline{
		\includegraphics[width=0.55\textwidth]{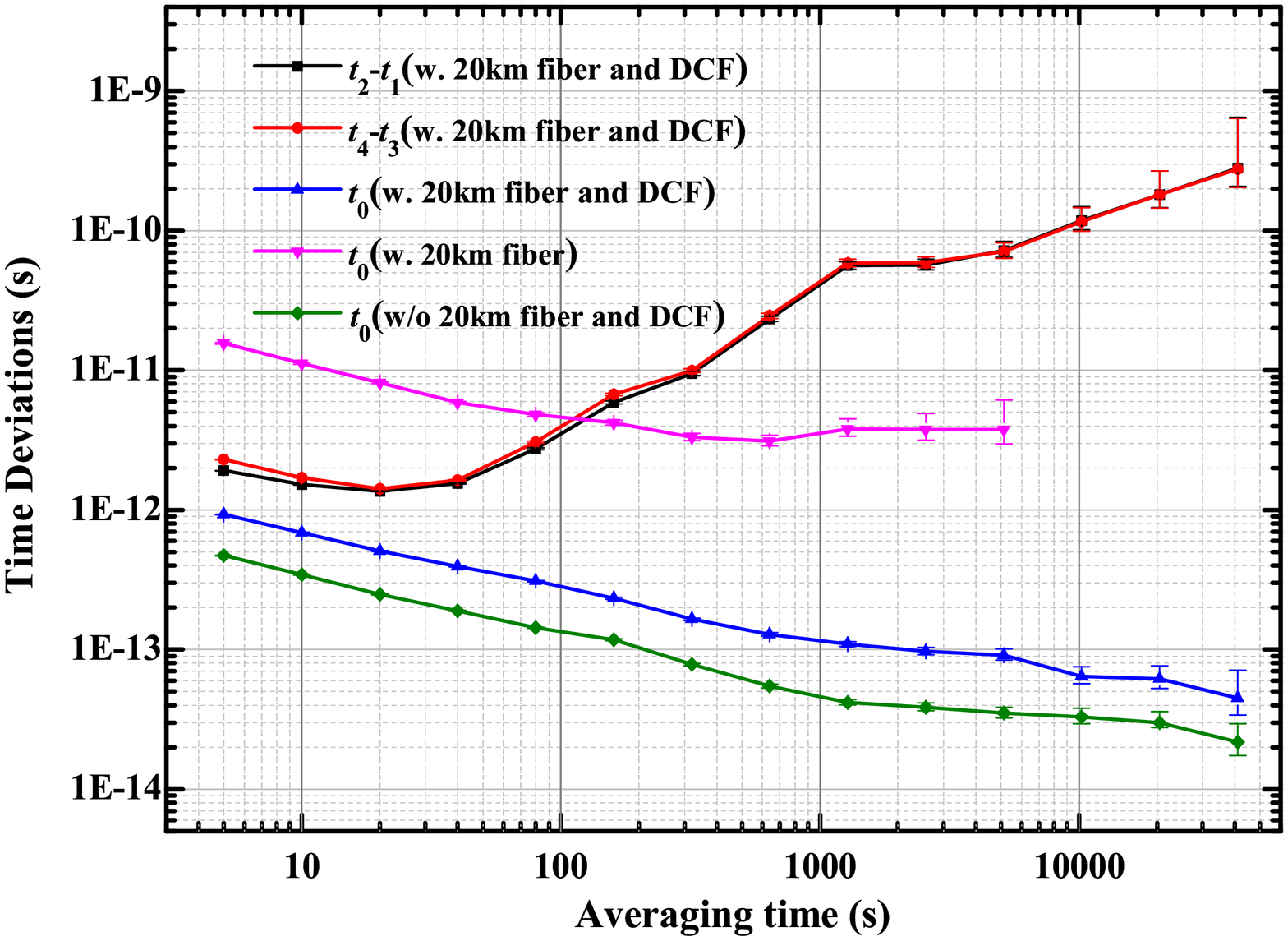}}
	\caption{\label{fig4} Measured TDEV of ${t_2} - {t_1}$, ${t_4} - {t_3}$  and ${t_0}$  between two clocks under the conditions that the $2.49$-km DCF was inserted, removed, and both the fiber coiling and the DCF were removed.}
\end{figure}

From the recovered temporal coincidence distributions, the averaged time differences ${t_2} - {t_1}$ , ${t_4} - {t_3}$, and ${t_0}$ over an interval of $5$ s can be extracted. The measured TDEV results with the DCF in place are presented in Fig.\ref{fig4}. As shown by black squares and red dots, the one-way differences ${t_2} - {t_1}$  and  ${t_4} - {t_3}$ fluctuate significantly, albeit with the same trend. By subtracting ${t_2} - {t_1}$ from ${t_4} - {t_3}$,  the transmission fluctuations can be canceled out; the results are shown by blue up-triangles. The TDEV of ${t_0}$ is $922$ fs over $5$-s averaging and a minimum of $45$ fs over $40960$-s averaging. With the DCF removed, the corresponding TDEV of ${t_0}$  is shown by magenta down-triangles. Over an averaging time of $5$ s, the TDEV is $15.6$ ps. At $5120$ s, the TDEV reaches a minimum of $3.77$ ps. The SDs of ${t_0}$ for these cases were measured to be $1.15$ ps and $17.35$ ps, respectively; these values agree well with those predicted in the theoretical simulation.

The performance of the system was tested by shortcutting the $20$-km fiber coiling with a $15$-m-long fiber and removing the DCF. The results, shown by olive diamonds in Fig.\ref{fig4}, set the lower limit for the achievable system stability. In this case, the FWHM of the coincidence distribution was measured to be $69.7$ ps, which corresponds to the combined FWHM jitters of the SNSPDs and ETs, with a total coincidence of about $2550$ over a $5$-s interval. The corresponding SD was measured to be $0.6$ ps, which agrees with the expected value of $0.59$ ps. Over an averaging time of $5$ s, the TDEV was $472$ fs; when the averaging time was extended to $40960$ s, the TDEV decreased to $21$ fs.

\begin{figure}
	\centerline{
		\includegraphics[width=0.55\textwidth]{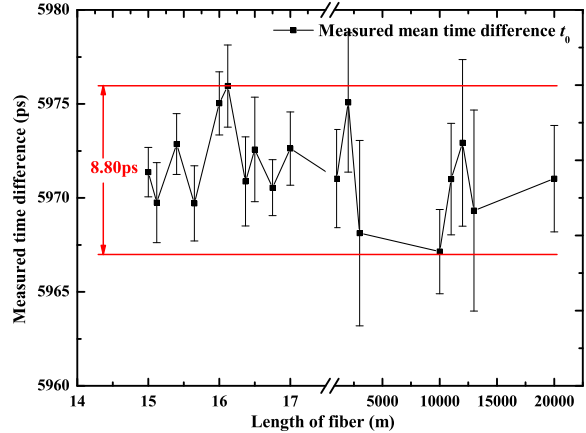}}
	\caption{\label{fig5} Measured mean value of time difference versus the inserted fiber length. }
\end{figure}
To investigate the accuracy of the setup in terms of its dependence on the length of the used fiber, we measured the mean time difference for fiber lengths varying from $15$ m to $20$ km. The obtained results are shown in Fig.\ref{fig5}, which shows a standard deviation of $2.46$ ps. According to Ref. \cite{lee2018}, the two-way quantum time transfer setup is robust against symmetric channel delay attacks because there is almost no correlation between the measured clock difference and the propagation distance ($ \le {\rm{0}}{\rm{.12}}$ps $km^{-1}$).

\section{Conclusion}

In summary, we have quantified and experimentally demonstrated a Q-TWTTOF scheme by using frequency-entangled pulses. Based on successful nonlocal time correlation measurement and nonlocal dispersion cancellation, a highly precise time transfer over a $20$-km fiber coiling with a stability of $922$ fs over $5$ s and $45$ fs over $40960$ s has been achieved. The lower limit for this achievable system stability was measured to be $472$ fs over $5$ s and $21$ fs over $40960$ s. The results could be further improved by using new ETs with sub-picosecond precision and higher acquisition rates, and applying new SNSPDs with lower timing jitters. The absolute time transfer accuracy as a function of the fiber length has been evaluated, and an uncertainty of $2.46$ ps in SD was found. Note that, in practical applications the DCF cannot be shared by the two distant parties. However, benefitted from the common mode noise suppression, as long as the two DCFs are set in the same environmental condition, the individual drifts induced by the two DCFs can be effectively depressed. If all common noise can be subtracted, a factor of $\sqrt {\rm{2}} $ increase is expected for the two-way long-term stabilities using two independent DCFs. In our experiment, the 45 fs TDEV at averaging time of 40960s can be expected to about 64 fs with two independent DCFs in the setup. However, the long-term stability may be deteriorated if the two DCFs are located remotely. Since the drift induced by the DCF is mainly due to its delay variation, which is dependent on the temperature variation and linearly proportional to the length \cite{sliwczynski2010}. Fiber Bragg Gratings (FBGs), which have much higher dispersion and therefore much shorter length, are to be used in the next experiment. As a comparison, we measured the performance of two FBGs in independent labs. The results show that performance of two FBGs was better than that of two DCFs. The details are shown in the appendix A. Besides improved time transfer stability over its classical counterpart (the comparison is shown in appendix B), this Q-TWTTOF system can also provide a secure distribution of timing information by incorporating with a Bell inequality test \cite{lee2014,Lamas-Linares2018}.

\begin{acknowledgments}
This work was supported by the National Natural Science Foundation of China (Grant Nos. 91336108, 11273024, 91636101, 91836301, 61801458, 61875205 and 61025023), the Research Equipment Development Project of the Chinese Academy of Sciences (Project Name: Quantum Improved Time Transfer Experiment System Based on Femtosecond Optical Frequency Combs), the National Youth Talent Support Program of China (Grant No.[2013] 33), the National Key R\&D Program of China (2017YFA0304000), and the Frontier Science Key Research Project of the Chinese Academy of Sciences (Grant Nos. QYZDB-SSW-SLH007 and QYZDB-SSW-JSC013).
\end{acknowledgments}

\appendix

\section{Discussion of the stability performance of different dispersive media against temperature fluctuations }
%

\begin{figure}
	\centerline{
		\includegraphics[width=0.55\textwidth]{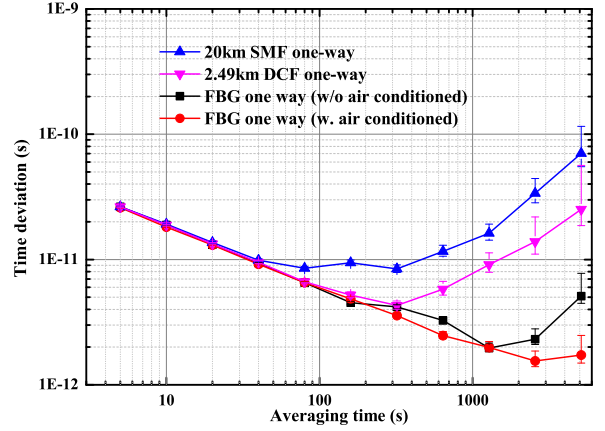}}
	\caption{\label{fig6} Comparison of one-way TDEVs in different cases.}
\end{figure}

To evaluate the stability of the delay induced by the 2.49 km DCF, we short connected the transmission path with a 15 m single mode fiber (SMF) and did the measurements for two different cases: 1) two 2.49 km DCFs were each connected into the idler arms at site A and B; 2) one 2.49 km DCF was shared by the idler arms at site A and B. Together with results of the case with 20 km SMF in the transmission path and no DCF in the idler arm, the one-way TDEVs are all shown in Fig.\ref{fig6}. Note that during the measurements, the coincidence rates were maintained about the same. The coincidence widths were also similar for the individual connections of 20 km SMF and 2.49 km DCF since the dispersions introduced by them were close.

From Fig.\ref{fig6}, one can see that the drift of the 2.49 km DCF (magenta down-triangles) is much smaller than that of the 20 km SMF (blue up-triangles). The results of DCF are 2.5 folds better than those of SMF over the averaging time of 1280 to 5120 s.

\begin{figure}
	\centerline{
		\includegraphics[width=0.55\textwidth]{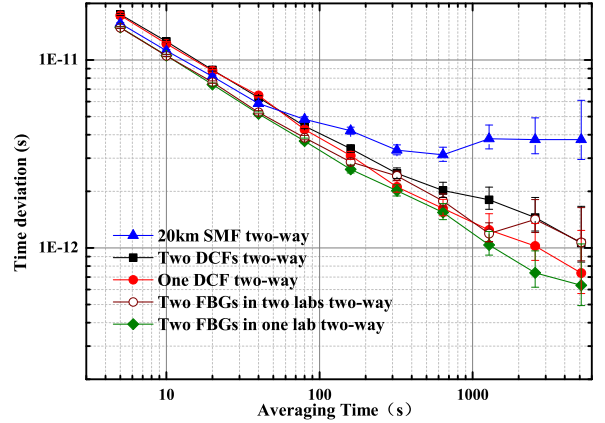}}
	\caption{\label{fig7} Comparison of two-way TDEVs in different cases.}
\end{figure}

By looking at the two-way results shown in Fig.\ref{fig7}, for the cases with only the 20 km SMF in the transmission path (blue up-triangles) and with one 2.49 km DCF shared by the idler arms of two sites A and B (red dots), most of the long-term deviations in the one-way setup are canceled based on the symmetric properties. 

Further comparing the two-way results for one DCF shared (red dots) and two DCFs used at each site (black squares), a $\sqrt {\rm{2}} $ ratio of the TDEV is observed over the averaging time of 1280 to 5120 s. This can be understood that two DCFs experience similar temperature variation and thus the common mode noise is subtracted.

Deterioration of the long-term stability would be expected if the two DCFs are located remotely, since the temperature variation experienced by them cannot be the same even when their environmental temperature can be well controlled. However, the Fiber Bragg Grating (FBG) can be a good solution because of its strong dispersion and very short length. Additionally, we measured the delay drift induced by two FBGs, the length of which is only about 6 m to compensate the dispersion of a 20 km fiber. To measure the delay drift induced by the FBG against temperature fluctuation, the scheme was built by replacing the DCFs in the idler arms with the two FBGs while the 20 km SMF being removed. Based on this scheme, we measured the time delay drifts of the FBGs located in two labs. Lab 1 was air conditioned with a periodic temperature variation range of 18.7 $^{\circ}$C to 20.3 $^{\circ}$C, while Lab 2 was air unconditioned with its temperature irregularly varying within a range of 17.1 $^{\circ}$C to 23.8 $^{\circ}$C. 

The corresponding TDEV results for the FBGs in two labs are plotted, as shown by red dots (lab 1) and black squares (lab 2) in Fig.\ref{fig6}. From it we can also see that, the FBG is highly temperature insensitive in comparison with the DCF and SMF.

Furthermore, we did the two-way TDEV measurements based on the above scheme. To give a clear demonstration, two more cases were tested. For case 1, both FBGs were placed in lab 1; for case 2, one FBG remained in lab 1 while the other was put in lab 2. The two-way TDEV results for the two cases are shown in Fig.\ref{fig7} together with those for using DCFs. We can see that, the two-way TDEV result for case 1 (olive diamonds) is even better than the previous scheme where one DCF was shared (red dots). The two-way TDEV for case 2 (wine hollow circles) is slightly worse than the case with one DCF shared, but is still better than that for two DCFs in the same lab (black squares). In future field experiments, we will replace DCF with FBG. 

\section{Comparison of Q-TWTTOF and TWTTOF}
\appendix


A quantitative comparison of Q-TWTTOF (quantum two-way time transfer over fiber) and TWTTOF ( two-way time transfer over fiber) is highly helpful to address the advantage of the quantum method. The traditional (TWTTOF) method is generally based on amplitude modulation on the optical carrier. To make a comparison with our result, a recent paper \cite{Liang2015} is taken as an example. On a 50 km-long fiber coiling, it reported a TWTTOF time stability of 6 ps/5s and 1.7 ps/100s. In our case, we used a fiber length of 20 km, and the achieved Q-TWTTOF time stabilities were measured to be 0.9ps/5s and 0.3ps/100s. According to our measurement, with the dispersion effect can be nonlocally canceled, the achievable time transfer stability can be only attributed to the experienced loss. Thus we can make a simple deduction of the time stability performance of our method on a 50 km fiber by just adding a virtual loss of 6 dB (0.2 dB/km*30 km). Under the same condition, the Q-TWTTOF time stabilities on a 50 km-long fiber coiling could be estimated to be 1.8 ps/5s and 0.6 ps/100s, which are better than the classical method. 

In spite of the great number of photons that give a significant advantage, there are many other factors that set the limit to the time stability of the traditional TWTTOF. Among them, the utilized terminal units in the TWTTOF scheme are key to the achievable stability performance. We note that researchers are also trying to adopt the component used in our quantum scheme to improve the time stability of the terminals. For example, as ETs (e.g., A033-ET from Eventech Ltd. has a 3.5-5 ps single-shot time resolution) can have better timing resolution than the time-interval counters (e.g., the best product SR620 from Stanford Research Systems has a 25 ps single-shot time resolution), they have been applied to TWTTOF as the timing signal receiver, and a TDEV better than 60 fs for averaging intervals from 100 s to 10 000 s was achieved after being calibrated for its temperature dependence to 100 fs/K \cite{Kodet2016}. For the time-of-arrival detection, single photon counting approach allows to reduce the systematic biases as much as possible, and is therefore adopted in the applications of Satellite Laser Ranging (SLR) and Time Transfer by Laser Link (T2L2). The long-term timing stability of the SPAD detectors combined with a sub-picosecond ET (NPET) was just recently evaluated, which gives a TDEV value of better than 100 fs. Further incorporating the normally used multi-mode Small Form Factor Pluggable (SFP) laser modules for generating the optical timing signal, the ultimate time transfer stability of the system has achieved a TDEV less than 1 ps for averaging times of hours \cite{Trojanek2018}. Compared with such classical time transfer system, our result shows a TDEV of 350 fs at 10 s averaging, and better than 30 fs for averaging time longer than 10000 s, which is also much better than the classical analogue.

%
%
%
%
%
%
%

\nocite{*}


\begin{thebibliography}{99}

\bibitem{krehlik2012}
P. Krehlik, {\L}. Sliwczynski, {\L}. Buczek, and M. Lipi{\'n}ski,
IEEE Transactions on Instrumentation and Measurement \textbf{61}, 10 (2012).

\bibitem{sliwczynski2013}
{\L}. {\'S}liwczy{\'n}ski, P. Krehlik, A. Czubla, {\L}. Buczek, and M. Lipi{\'n}ski, 
Metrologia \textbf{50}, 133 (2013).

\bibitem{lopez2013}
O. Lopez, A. Kanj, P.-E. Pottie, D. Rovera, J. Achkar, C. Chardonnet, A. Amy-Klein, and G. Santarelli, 
Applied Physics B \textbf{110}, 1 (2013).

\bibitem{Rabindran2008}
J. B. Rabindran, S. Y. Kim, K. V. Bineesh, and D. W. Park, 
Metrologia \textbf{45}, 2 (2008)

\bibitem{Jiang2017}
Z. Jiang, Y. Huan, V. Zhang, and P. Dirk, 
BIPM Technical Memorandum, TM268 V2a \textbf (2017).

\bibitem{piester2009}
D. Piester, M. Fujieda, M. Rost, and A. Bauch, 
41st PTTI, Santa Ana Pueblo, New Mexico \textbf(2009).

\bibitem{rost2012}
M. Rost, D. Piester, W. Yang, T. Feldmann, T. Wübbena, and A. Bauch, 
Metrologia \textbf{49}, 6 (2012).

\bibitem{lopez2015}
O. Lopez , F. K{\'e}f{\'e}lian, H. Jiang, A. Haboucha, A. Bercy, F. Stefani, B. Chanteau, A. Kanj, D. Rovera, J. Achkar, C. Chardonnet, P. Pottie, A. Amy-Klein, and G. Santarelli, 
Comptes Rendus Physique \textbf{16}, 5 (2015).

\bibitem{wang2012}
B. Wang, C. Gao, W. Chen, J. Miao, X. Zhu, Y. Bai, J. Zhang, Y. Feng, T. Li, and L. Wang, 
Sci. Rep. \textbf{2}, 556 (2012).

\bibitem{yin2014}
F. Yin, Z. Wu, Y. Dai, T. Ren, K. Xu, J. Lin, and G. Tang, 
Opt. Lett. \textbf{39}, 10 (2014).

\bibitem{raupach2014}
S. M. Raupach and G. Grosche, 
IEEE Trans. Ultrason. Ferroelectr. Freq. Control \textbf{61}, 6 (2014).

\bibitem{chen2015}
X. Chen, J. Lu, Y. Cui, J. Zhang, X. Lu, X. Tian, C. Ci, B. Liu, H. Wu, T. Tang, K. Shi, and Z. Zhang, 
Sci. Rep. \textbf{5}, 18343 (2015).

\bibitem{krehlik2017}
P. Krehlik, H. Schnatz, and {\L}.{\'S}liwczy{\'n}ski, 
IEEE Trans. Ultrason. Ferroelectr. Freq. Control  \textbf{64}, 1884 (2017).

\bibitem{lessing2017}
M. Lessing, H. Margolis, C. T. A. Brown, and G. Marra, 
Applied Physics Letters. \textbf{110}, 22 (2017).

\bibitem{peng2013}
M. Y. Peng, P. T. Callahan, A. H. Nejadmalayeri, S. Valente, M. Xin, L. Gr{\"u}ner-Nielsen, E. M. Monberg, M. Yan, J. M. Fini, and F. X. K{\"a}rtner, 
Optics Express, \textbf{21}, 17 (2013)

\bibitem{xin2014}
M. Xin, K. {\c{S}}afak, M. Y. Peng, P. T. Callahan, and F. X. K{\"a}rtner, 
Optics express \textbf{22}, 12 (2014)

\bibitem{ning2014}
B. Ning, S. Y. Zhang, D. Hou, J. T. Wu, Z. B. Li, and J. Y. Zhao, 
Sci. Rep. \textbf{4}, 5109 (2014).



\bibitem{jafarnia2012}
A. Jafarnia-Jahromi, A. Broumandan, J. Nielsen, and G. Lachapelle, 
International Journal of Navigation and Observationn \textbf{12},(2012)

\bibitem{angel2014}
J.J. Angel,
Financial Review \textbf{2}, 49 (2014)

\bibitem{narula2018}
L. Narula, and T.E. Humphreys, 
IEEE Journal of Selected Topics in Signal Processing \textbf{12}, 4 (2018).

\bibitem{exertier2014}
P. Exertier, E. Samain, N. Martin, C. Courde, M. Laas-Bourez, C. Foussard, and Ph. Guillemot, 
Advances in Space Research \textbf{54}, 11 (2014).

\bibitem{samain2015}
E. Samain, P. Exertier, C. Courde, P. Fridelance, P. Guillemot, M. Laas-Bourez, and J-M Torre, 
Metrologia \textbf{52}, 2 (2015).


\bibitem{giovannetti2011}
V. Giovannetti, S. Lloyd, and L. Maccone, 
Nat. Pho. \textbf{5}, 4 (2011).

\bibitem{giovannetti2001a}
V. Giovannetti, S. Lloyd, and L. Maccone, 
Nature \textbf{412}, 417 (2001).

\bibitem{giovannetti2001b}
V. Giovannetti, S. Lloyd, and L. Maccone, 
Phys. Rev. A \textbf{65}, 2 (2001).

\bibitem{bahder2004}
T. B. Bahder and W. M. Golding, 
AIP Conference Proceedings \textbf{734}, 1 (2004).

\bibitem{wang2016}
J. Wang, Z. Tian, J. Jing, and H. Fan, 
Phys. Rev. D \textbf{93}, 6 (2016).

\bibitem{valencia2004}
A. Valencia, G. Scarcelli, and Y. Shih, 
Applied Physics Letters \textbf{85}, 13 (2004).

\bibitem{quan2016}
R. Quan, Y. Zhai, M. Wang, F. Hou, S. Wang, X. Xiang, T. Liu, S. Zhang, and R. Dong, 
Sci. Rep. \textbf{6}, 30453 (2016).

\bibitem{quan2019}
R. Quan, R. Dong, Y. Zhai, F. Hou, X. Xiang, H. Zhou, C. Lv, Z. Wang, L, You, T. Liu, and S. Zhang, 
Optics Letters \textbf{44}, 3 (2019).

\bibitem{giovannetti2002}
V. Giovannetti, S. Lloyd, and L. Maccone, 
Journal of Optics B: Quantum and Semiclassical Optics \textbf{4}, 4 (2002).

\bibitem{lee2014}
C. Lee, Z. Zhang, G. R. Steinbrecher, H. Zhou, J. Mower, T. Zhong, L. Wang, X. Hu, R. D. Horansky, V. B. Verma, A. E. Lita, R. P. Mirin, F. Marsili, M. D. Shaw, S. W. Nam, G. W. Wornell, F. N. C. Wong, J. H. Shapiro,  and D. Englund 
Phys. Rev. A \textbf{90}, 062331 (2014).

\bibitem{Lamas-Linares2018}
A. Lamas-Linares and J. Troupe, 
Advances in Photonics of Quantum Computing, Memory, and Communication XI. International Society for Optics and Photonics \textbf{10547}, (2018).

\bibitem{franson1992}
J. D. Franson, 
Phys. Rev. A \textbf{45}, 5 (1992).

\bibitem{baek2009}
S. Y. Baek, Y. W. Cho, and Y. H. Kim, 
Optics Express \textbf{17}, 21 (2009).

\bibitem{O2011}
K. A. O'Donnell, 
Phys. Rev.lett. \textbf{106}, 6 (2011).

\bibitem{maclean2018}
J.P.W. MacLean, J.M. Donohue,  and K.J. Resch,
Phys. Rev. Lett. \textbf{120}, 5 (2018)

\bibitem{Panek2013}
P. Panek, J. Kodet, and I. Prochazka, 
European Frequency and Time Forum  International Frequency Control Symposium (2013).

\bibitem{Wu2017}
J. Wu, L. You, S. Chen, H. Li, Y. He, C. Lv, Z. Wang and X. Xie, 
Applied Optics \textbf{56}, 8 (2017).

\bibitem{lee2018}
J. Lee, L.Shen, A. Cerè, J. Troupe, A. Lamas-Linares, and C.  Kurtsiefer, 
arXiv:1812.08450 (2018).

\bibitem{quan}
R. Quan, R. Dong, F. Hou, T. Liu, and S. Zhang, 
arXiv:1907.08925.

\bibitem{glauber1963}
R. J. Glauber, 
Physical Review \textbf{130}, 6 (1963).

\bibitem{Hou2016}
F. Hou, X. Xiang, R. Quan, M. Wang, Y. Zhai, S. Wang, T. Liu, S. Zhang, and R. Dong, 
Applied Physics B \textbf{122}, 5 (2016).


\bibitem{sliwczynski2010}
{\L}. {\'S}liwczy{\'n}ski, P. Krehlik, and M. Lipi{\'n}ski,
Measurement Science Technology \textbf{21}, 7 (2010).

\bibitem{Liang2015}
K. Liang, A. Zhang, Z. Yang, W. Chen, W. Wang, 
Frequency Control Symposium \& the European Frequency \& Time Forum. IEEE, 2015.

\bibitem{Kodet2016}
J. Kodet, P. P{\'a}nek, I. Proch{\'a}zka, 
Metrologia \textbf{53},1 2016.

\bibitem{Trojanek2018}
P. Trojaneka and I. Prochazka, 
Review of Scientific Instruments \textbf{89},8 2018.




\end{thebibliography}

\end{document}